# Heuristic for Network Coverage Optimization Applied in Finding Organizational Change Agents

Błażej Żak[1], Anita Zbieg[2]

[1] Faculty of Computer Science and Mgmt., Wroclaw University of Technology
[2] Institute of Psychology, University of Wroclaw; Wroclaw University of Economics

blazej.zak@pwr.wroc.pl, anita.zbieg@ue.wroc.pl

*Abstract* - **Authors compare different ways of selecting change agents within network analysis paradigm and propose a new algorithm of doing so. All methods are evaluated against network coverage measure that calculates how many network members can be directly reached by selected nodes. Results from the analysis of organizational network show that compared to other methods the proposed algorithm provides better network coverage, at the same time selecting change agents that are well connected, influential opinion leaders.**

*Keywords- sna, ona, organizational network analysis, organizational change, change agents, network coverage*

## INTRODUCTION

The majority of companies faces a problem of how to manage an organizational change [1] [2] [3] [4]. Conducting the change can be a difficult task because of a number of forces within and outside an organization that make it uncontrollable and unpredictable [5]. Social network analysis (SNA) [6] [7] can be helpful, especially in finding people that can make change implementation effective and successful by spreading information as well as providing valuable feedback [7]. Such people are called *opinion leaders* or *change agents* and once they are engaged they act as catalysts for change [2] [7] [9] [10] [11]. Change agents can be considered as people able to reach many others within a company, inform them about a change and teach others to successfully deal with it in practice [11] [12]. Change agents can be also recruited from influential employees called opinion leaders that are able to influence other people attitudes towards a change [2] [12].

The paper proposes a simple heuristic algorithm of finding change agents by analysis of a company's organizational network. The paper aims to focus on *network coverage* which is the extent to which change agents directly reach others within the network. Considering it important and not already applied to finding change agents within an organization, the paper proposes network coverage as a main criterion in evaluation of change agents selection process, both related to existing measures and the proposed algorithm.

## BACKGROUND

### A. Finding change agents with traditional centrality measures

Basic tools in social network analysis perspective include network centrality measures [7] [13] [14] helpful in finding people that plays key roles in company's social network.

*1) Degree centrality*

It is calculated as a number of ties of a node that connect it directly with others in the network.

*In-degree centrality* - is calculated as a number of node's incoming ties. Employees with a high in-degree measure tend to be an authority. They have many indications received from others because of some resource (e.g. information, knowledge, power) they possess.

*Out-degree centrality* - is calculated as a number of node's outgoing ties. Employees with a high out-degree centrality are active networkers, but it's not sure that their voice are heard by others, hence the measure is rather not applied in a process of change agents selection.

*2) Betweenness centrality*

It is calculated as a number of shortest paths within the network that pass by a node. Employees with a high betweenness serve as bridges connecting diverse groups.

*3) Closeness centrality*

It is calculated for a given node as an inverse of an average number of steps (edges) to all other nodes in a network and reflects the distance to others. Employees with a high closeness tend to have a current access to information flowing within a network.

*4) Eigenvector centrality*

It is calculated as the number of connections of a given node with a particular attention to those with many links to others also evaluated according to how many links have their neighbors... etc.. Employees with high eigenvector centrality tend to have the global impact in the network.

## THEORETICAL FRAMEWORK

### A. Algorithm for finding high coverage nodes

The algorithm presented below provides coverage of the network and works as follows: in each step we add to selected nodes collection a node that contributes highest number of covered nodes. We repeat this step until total coverage of selected nodes is higher than desired. Therefore this approach might be described as a greedy algorithm [15] for selecting high coverage nodes in a given graph. As a node coverage we consider all other nodes connected to the given node by incoming relationships, and the node itself. The algorithm might not be optimal but should work well as a heuristics. However optimal algorithm would require us to check all possible combinations of nodes what would be too computationally intensive and hence infeasible in a real world. The problem of finding minimal set of high coverage nodes in a network for computational point of view is very similar to NP-hard knapstack problem [16]. Therefore other heuristics developed for the knapstack problem might be used to develop more optimal solutions. Other algorithms have been also already proposed [17] [18].

It should be also noticed that the algorithm works on nodes' in-degree. We found two reasons for choosing in-degree as a criterion for the nodes' selection. First, one can potentially increase the chance that among chosen change agents are employees with network authority. Secondly, employee's in-degree is built on indications or interactions made by others. That makes it more objective way of node's evaluation than other centrality measures that depend also on its own indications. Using in-degree measure can also help to avoid the main problem that accompanies network analysis based on sociometric surveys [19] which is the lack of data.

**ALGORITHM 1** – ALGORITHM FINDING HIGH COVERAGE NODES FOR A GIVEN GRAPH

**Function** FindHighCoverageNodes(nodes, edges, target_coverage)
1. Calculate in-degree for each node in the graph
2. Sort graph nodes according to in degree descending
3. While network coverage < target_coverage do
    a. Go through sorted list of unselected nodes and calculate their contribution (uncommon nodes) to overall coverage. Stop when max(coverage contribution) > current node in-degree
    b. Add node contributing highest coverage to selected nodes collection
    c. Calculate network coverage of all selected nodes
4. Return selected nodes

### B. Research questions

Direct reach of the majority of company employees made by change agents is an important factor of a large scale organizational change success. First of all, selected change agents should be able to broadly and directly reach the organizational network. To verify this assumption, we follow the Pareto principle or 80/20 which states that roughly 80% of the effects come from 20% of the causes [20]. But change agents should be still more influential or important within the network than an average node what builds the second criterion of method evaluation. Thus, raise the questions motivating the research: (1) *To what extent is the proposed algorithm (finding high coverage nodes for a given graph) an efficient way of selecting nodes covering the network* (2) *Is the proposed algorithm more efficient way of selecting nodes covering the network than SNA centrality measures?* and (3) *To what extent does a list of nodes selected by the algorithm correspond to the list of nodes selected by SNA centrality measures?*

## STUDY AND DATA ANALYSIS

To test if the network coverage is both, effectively obtained by proposed method (first research question), and obtained more effectively with an algorithm than by methods based on centrality measures (second research question), and finally verify the quality of selected nodes (third research question), the authors analyzed an intra-organizational network of composed of 215 employees.

### A. Data gathering

The authors investigated an intra-organizational network of a production and sales company built of the relations of collaboration between 215 employees. The study was conducted in a two week period of march 2014 and involved all company desk employees (*N*=215).

In a sociometric survey, the authors used traditional egocentric questions [19] popular in organizational studies [13] [21] [22] [23]. With a cloud based, participatory network mapping platform for organizational network studies [24], respondents were asked the question: *With whom do you directly work to perform your everyday tasks?* and write in response the names of their collaborators. The survey was completed by 142 employees. Thus, the overall response rate was 66% which is a result methodologically acceptable and similar to the results obtained in other studies [22].

### B. Intra-organizational network description

The obtained intra-organizational social network presents the collaboration network between a company employees. The network is directed and it is characterized by the following general properties: (1) the number of nodes: 215 (even if not all employees completed the survey, they've been indicated by others); (2) the number of edges: 2225, (3) the network density was 4,8%; (3) average node degree: 20,69.

The employees in our sample were characterized by: (1) the following distribution of hierarchy position: directors 8,6%, managers 29,4%, specialists 62%; (2) tenure: 17,6% less than a year, 35,3% more than a year and less than 5 years, 17,6% more than 5 years and less than 10 years and 29,5% more than 10 years.

*C. Evaluation of network coverage*

To answer the research questions and evaluate created method of network coverage evaluation, we have followed the three step plan of analysis.

*1) Network coverage effectiveness calculations*

First, we've calculated the centrality measures for each network node: degree, in-degree, closeness, betweenness and eigenvector centrality. Secondly, we've started analyzing the coverage of the network according to created algorithm. Once, we reached the network coverage close to 80%, we made the rank of selected nodes. Next, we've calculated the ranks of nodes this time selected by centrality measures according to the rule: the higher centrality measure possesses a node, the higher its position on the rank. Finally, for nodes selected according to centrality measures has been calculated the network coverage indicator.

Having ranks created with both, created algorithm and classical centrality measures the results of network coverage obtained at each ranking level have been compared. It is presented in Table 1. The network coverage indicator for all methods varies from 77% to 83% for 50 selected nodes. The smallest coverage is obtained by selecting nodes by eigenvector centrality rank (77%), and better result give: closeness centrality rank (79%), betweenness rank (80%) and in-degree rank (81%).

Created method was the most effective way of selecting nodes for network covering (83%) and the difference in number of nodes reached by proposed method is 178 while selecting nodes with the less effective way is 165. If this difference doesn't seem huge while calculating the network of 215 nodes, it should grow for bigger networks. Moreover, the difference between network coverage calculated with diverse ranks achieved similar results for each level of ranking (1, 2, 3 nodes ecc.) what suggests that the advantage of an algorithm is fairly stable.

Once, we reached the network coverage close to 80%, we counted the number of nodes in the rank. As shows the same table, the Pareto 20/80 rule surprisingly has been maintained for the network coverage obtained with all methods. Fifty nodes selected for about 80% of network coverage results as 23% of nodes present in the analyzed network (*N*=215). That suggests that the 20/80 rule can be used in network coverage procedures, but to generalize obtained results it should be verified in future works by analyzing more networks.

TABLE 1
NETWORK COVERAGE FOR CHANGE AGENTS SELECTED BY DIFFERENT METHODS:

| Number of employees selected | in-degree rank | betweenness rank | closeness rank | eigenvector rank | suggested algorithm |
|---|---|---|---|---|---|
| 1 | 17% | 13% | 12% | 10% | **17%** |
| 2 | 22% | 20% | 19% | 19% | **28%** |
| 3 | 32% | 24% | 25% | 23% | **34%** |
| 4 | 35% | 30% | 30% | 25% | **39%** |
| 5 | 38% | 32% | 31% | 29% | **43%** |
| 10 | 50% | 45% | 42% | 44% | **56%** |
| 20 | 59% | 63% | 58% | 57% | **69%** |
| 30 | 66% | 70% | 67% | 64% | **74%** |
| 40 | 75% | 75% | 74% | 70% | **79%** |
| 50 | 81% | 80% | 79% | 77% | **83%** |

*2) Verification of the quality of selected nodes*

Once we knew that the high network coverage (about 80%) is most effectively obtained with the proposed algorithm, we wanted to verify the quality of selected nodes involved in the process of network coverage. While network coverage is an important factor of the change implementation process, network position of selected change agents is also important. To do so, we tested the strength of association between the position of node in a rank created by an algorithm with the position of node in a rank created according to the in-degree, betweenness, closeness and eigenvector centrality measures. The association between ranks was tested with the Spearman's rank-order correlation, as it has already been applied in similar studies [22] [25].

TABLE 2
THE STRENGTH OF ASSOCIATION BETWEEN NODE RANK
SELECTED BY SUGGESTED ALGORITHM AND CENTRALITY MEASURES

|  | in-degree centrality | betweenness centrality | closeness centrality | eigenvector centrality |
|---|---|---|---|---|
| Suggested algorithm coverage rank decreasing | .921 | .834 | .782 | .689 |

Spearman's rank-order correlation ($r_s$), p= 0.01 (two tailed)

The results presents Table 2. All centrality measures highly correlate with the position of a node in the rank created with an algorithm. As expected, the highest correlation results with in-degree because an algorithm operates on this measure incoming relationships. Nonetheless, other centrality measures also well describe the nodes present in a rank. That suggests that all nodes selected as change agents not only can significantly cover the network but are also important nodes described as centrally positioned within the studied network.

## CONCLUSION

Authors proposed heuristic for network coverage optimization and presented it as a way of finding change agents. Nodes selected with an algorithm can directly reach the network broader than nodes selected with traditional centrality measures being in the same time characterized with high centrality in terms of network bonding, authority, bridging, pulse and influence.